\documentstyle[12pt]{article}

\begin{document}
\begin{titlepage}

\title{Signature change induces compactification
         \thanks{
         Work supported by the Austrian Academy of Sciences
         in the framework of the ''Austrian Programme for
         Advanced Research and Technology''.}}

\author{Franz Embacher\\
        Institut f\"ur Theoretische Physik\\
        Universit\"at Wien\\
        Boltzmanngasse 5\\
        A-1090 Wien\\
        \\
        E-mail: fe@pap.univie.ac.at\\
       \\
       UWThPh-1994-47\\
       gr-qc/9410012
        }
\date{}

\maketitle

\begin{abstract}
It is shown -- using a FRW model with ${\bf S}^3 \times {\bf S}^6$
as spatial sections and a positive cosmological constant -- that
classical signature change implies a new compactification
mechanism. The internal scale factor is of the order
$\Lambda^{-1/2}$, and the solutions are stable against
small perturbatons. In the case of compactified ${\bf S}^6$, it is
shown that the effective four-dimensional space-time metric has
Lorentzian signature, undergoes exponential inflation in
${\bf S}^3$ and is unique. Speculations concerning relations
to quantum cosmology and conceivable modifications are added.
\end{abstract}

\end{titlepage}

\section{Introduction}

The aim of this paper is to present a mechanism that combines the
issues of classical signature change and compactification. Both
topics are of general interest in the recent discussion on
gravity theory and cosmology, the former being comparably
young, whereas the latter may be called an old
tradition.
\medskip

Our approach is mainly a cosmological one. Given a higher
dimensional model including gravity, and assuming the
spatial sections to be of topology, say
${\bf S}^3 \times {\cal M}$,
one would like to know whether there is a dynamical mechanism
that drags the size of ${\cal M}$ down and keeps it
in a stable way at an unobservably small (''compactified'')
scale. Without repeating all the advantages that the
existence of internal dimensions could provide for various
aspects of particle physics and field theory
\cite{BailinLove}, we just
mention that the Kaluzu-Klein idea is crucial for several
theories (like supergravity and superstring models) whose
formulation requires a particular number of dimensions.
\medskip

In the cosmological context, there are a lot of mechanisms at
hand that prevent some of the dimensions from
expanding \cite{BailinLove}--\cite{cos}.
The long-time stability of the geometrical configurations
emerging is a more subtle problem, and the most popular
methods to prevent internal dimensions from
collapsing use particular forms of matter couplings. One
famous example is the Freund-Rubin compactification
in eleven-dimensional supergravity
\cite{FreundRubin}. Among various other
possibilities, we just mention the gravitational Casimir
effect \cite{casimir}
and the approaches provided by quantum cosmology
\cite{qucos1}--\cite{qucos2}.
\medskip

Our motivation lies in the search for fundamental
compactifications mechanisms, i.e. models that are mainly
based on gravity. Leaving apart sophisticated matter
couplings, we choose the very simple model of pure
gravity with a positive cosmological constant.
\medskip

The second issue we have mentioned is classical signature
change. In this approach, one allows the metric to be
of different signature in different regions of the total
manifold considered. There has been a recent discussion
on the relevance and the physical nature of such models
\cite{EllisSumeruketal}--\cite{FE4}.
In the most important version, the signature of the metric
may change from Euclidean to Lorentzian type. Usually, the
Euclidean domain is considered as related to the early
universe, and the existence of a physical time is
a consequence of a signature change
\cite{EllisSumeruketal}. In some sense,
this approach is a classical alternative to quantum
cosmology, where Euclidean and Lorentzian geometries
are interlinked as well, but in a different way
\cite{HartleHawking}.
We will use a specific approach to classical signature change
that is effectively based on the junction conditions
advocated in Ref. \cite{EllisSumeruketal} (and that has
been denoted {\it weak} signature change in Ref. \cite{FE4}).
\medskip

Putting these two ingredients together, we can show that
that a repeated sequence of signature changes
can effectively compactify and stabilize internal spaces.
This has been worked out in detail for a
Friedmann-Robertson-Walker (FRW)
model with ${\bf S}^3 \times {\bf S}^6$
as spatial sections. Some of the ideas contained in this
paper have been developed in Ref. \cite{FE1}.
\medskip

Our presentation begins with an outline
of the model and the choice of convenient variables in
Sect.2. The compactification mechanism is discussed in
Sect.3. It is based on the interplay between the causal
structure of the Wheeler-DeWitt metric and the sign of the
potential $W$ showing up in the equations of motion of
the two scale factors. Euclidean solutions have the tendency
to leave the domain $W<0$, whereas a large portion of
Lorentzian solutions cannot escape to arbitrarily
large values of $W$, but recollapse. Thus, assuming
actual occurence of a signature change
whenever it is possible
(the condition being $W=0$), one obtains a family
of metrics whose signature type ''oscillates'' between
Lorentzian and Euclidean, and whose location in
the minisuperspace built up by the scale factors is
near the curve $W=0$ (which in turn describes
compactification of either sphere). An
example for such a metric that has been obtained
numerically, is displayed.
\medskip

In Sect.4 we assume that the effective (physical) metric
is provided by a sort of coarse graining average over
the true (oscillating) one. For the case of compactified
${\bf S}^6$ we show that the resulting effective metric is
unique, has Lorentzian signature (which is due to a small
domination of the Lorentzian periods over the
Euclidean ones) and describes exponential inflation
of the remaining three-space ${\bf S}^3$.
Sect.5 is devoted to the discussion of physical
problems related to quantum gravity and the necessity to
exit inflation. Various conceivable modifications
and generalizations are pointed out. In the last Section
we comment on the alternative case of compactified
${\bf S}^3$, and on the generalization to a
${\bf S}^m \times {\bf S}^n$ model. The structure of
the equations of motion turns out to be such that only
internal spaces with non-vanishing Ricci-curvature can
compactify by signature change.
\medskip

\section{The model and its variables}
\setcounter{equation}{0}

We consider a FRW model with the
product  ${\bf S}^3 \times {\bf S}^6$ of round spheres as
spatial sections of space-time, and a positive cosmological
constant $\Lambda$. Some of the results we will achieve may be
generalized to ${\bf S}^m \times {\bf S}^n$, and we comment
on this at the end of the paper. The class of metrics is thus
described by
\begin{equation}
ds^2 = \mp \, {\cal N}(t)^2 dt^2 + a_1(t)^2 d\sigma_3^2 +
       a_2(t)^2 d\sigma_6^2,
\label{2.1}
\end{equation}
where $d\sigma_n^2$ is the line-element on the round unit n-sphere.
Here and in what follows, the upper sign belongs to the Lorentzian
(i.e. Pseudo-Riemannian) and the lower sign to the Euclidean
(Riemannian) version. In this Section, we consider these cases
independently. The ten-dimensional Einstein-Hilbert action
including a positive cosmological constant and the usual
boundary term that absorbs the second time-derivatives, may be
written as
\begin{equation}
S = \mp \, C \int_{{\cal M}_{10}} d^{\,10}x \sqrt{|g|}
    (\,{}^{10}R + 2\, \Lambda)
    \mp\, 2\, C \int_{\partial {\cal M}_{10}} d^{\,9} x \sqrt{h} K,
\label{2.2}
\end{equation}
where
\begin{equation}
C = \frac{m_P^2}{16 \pi} (\rm{volume\, of\, internal\, space\, today})^{-1},
\label{2.3}
\end{equation}
$h_{ij}$ the metric induced by $g_{\mu\nu}$ on the boundary
$\partial {\cal M}_{10}$
and $K$ the trace of its extrinsic curvature.
The above choice of $C$ ensures the correct gravitational
constant today if ${\bf S}^6$ is compactified (as internal space)
at a small value of its
scale factor $a_2$. The ${\bf S}^3$ is then the physical (external)
space we observe. (However, since we are interested here only in
classical solutions to Einstein's field equations, the constant $C$,
being just a prefactor of the action, drops out anyway). Inserting
(\ref{2.1}) into (\ref{2.2}) results into
\begin{eqnarray}
S = w C \int dt \, \bigg(
 \frac{a_1^3 a_2^6}{\cal N}
  \Big( - 6\, \frac{\dot{a_1}^2}{a_1^2}
    - 36 \,\frac{\dot{a_1}}{a_1}\frac{\dot{a_2}}{a_2}
    - 30 \,\frac{\dot{a_2}^2}{a_2^2} \Big)
             \nonumber\\
 \mp \,{\cal N} \Big( - 6\, a_1 a_2^6 -30\, a_1^3 a_2^4
                        + 2 \Lambda a_1^3 a_2^6 \Big)
          \bigg). \label{2.4}\\
\nonumber
\end{eqnarray}
Here, $w=32 \pi^5/15$, and the volume of the $t = const$ space sections
is given by
\begin{equation}
V = \int_{{\cal M}_9}d^{\,9} x\sqrt{h} = w a_1^3 a_2^6.
\label{2.6}
\end{equation}
As is well known, the variation of this action with respect to
the variables ${\cal N}(t)$, $a_1(t)$ and $a_2(t)$ yields the
full set of Einstein's field equations for the ansatz (\ref{2.1}).
\medskip

In order to diagonalize the kinetic part of the action, we change
variables according to \cite{FE1}
\begin{eqnarray}
N & = & \frac{\Lambda}{3 \sqrt{10}}\, a_2\, {\cal N},\label{2.13/2}\\
u & = & \frac{\Lambda^2}{180}\, a_1^2\, a_2^2,\label{2.7}\\
v & = & \frac{\Lambda^3}{5400 \sqrt{5}}\, a_1 \,a_2^5,\label{2.15/2}\\
\nonumber
\end{eqnarray}
and define (using $\Lambda \equiv \ell_{\Lambda}^{-2}$
and $m_P \equiv \ell_{P}^{-1}$)
\begin{equation}
\Lambda_{\rm {eff}} = \frac{375^{1/4} \Lambda}{2^{1/8}\,6\, (w C)^{1/4}}
                   \equiv \Bigg(
                   \frac{1}{\pi}
                   \bigg(\frac{5}{3\sqrt{2}}\bigg)^3
                   \bigg (\frac{\ell_P}{\ell_{\Lambda}}\bigg)^2
                   \bigg(\frac{a_2(\rm{today})}{\ell_{\Lambda}}\bigg)^6
                                        \Bigg)^{1/4}.
\label{2.8}
\end{equation}
The action (\ref{2.4}) simplifies to
\begin{equation}
S = \frac{1}{\Lambda_{\rm{eff}}^4}
   \int dt \left( -\,\frac{\dot u \dot v}{N}
                 \mp \, N W(u,v) \right).
\label{2.9}
\end{equation}
The potential is given by
\begin{equation}
W(u,v) = - v - u^{3/2} + 2\, u^{5/4} v^{1/2},
\label{2.10}
\end{equation}
the origin of the three contributions being the curvature of ${\bf S}^3$,
the curvature of ${\bf S}^6$ and the cosmological constant, respectively.
The lapse variable $N$ corresponds to the choice of a particular
(Lorentzian or Euclidean)
''time'' parameter $t$, and the only dynamical degrees of freedom
in our model are $u$ and $v$, both ranging from 0 to $\infty$. A ''point''
$(u,v)$ in this minisuperspace represents a nine-geometry. Classical
solutions may be displayed as trajectories (a continuous sequence
of nine-geometries), and a (gauge) choice fixing the time parameter $t$
leads to parametrized curves $(u(t),v(t))$. The equations of motion
following from variation of (\ref{2.9}) are
\begin{eqnarray}
\frac{\dot{u}\dot{v}}{N^2} & = & \pm \, W,\label{constr}\\
\frac{1}{N}\frac{d}{dt}\bigg(\frac{\dot{u}}{N}\bigg) & =
            & \pm \, \partial_v W,\label{2.11}\\
\frac{1}{N}\frac{d}{dt}\bigg(\frac{\dot{v}}{N}\bigg) & =
            & \pm \, \partial_u W,\label{2.23/2}\\
\nonumber
\end{eqnarray}
the first being the constraint due to the absence of $\dot{N}$ in the
action. It translates to the Wheeler-DeWitt equation in the framework
of quantum coslomogy and, once being satisfied at $t=t_0$, it is
conserved in time by the two remaining equations.
\medskip

The kinetic part of the action (\ref{2.9}) shows that $u$ and $v$ are
''lightlike'' coordinates with respect to the Wheeler-DeWitt metric
$ds^2_{WDW} = - du dv$. The constraint equation (\ref{constr})
implies that
Lorentzian trajectories that represent classical solutions are timelike
($du dv>0$) curves in the regions of minisuperspace where $W>0$ and
spacelike ($du dv<0$) when $W<0$. Conversely, Euclidean trajectories
are spacelike if $W>0$ and timelike if $W<0$.
\medskip

One may perform further transformations of variables
$\overline u=\overline u(u),\,\overline v=\overline v(v)$,
thereby retaining the structures
(\ref{2.9}) and (\ref{constr}-\ref{2.23/2}) if one sets
\begin{equation}
\overline W=W \frac{du}{d\,\overline u}
\frac{dv}{d\,\overline v}, \qquad
\overline N=N \frac{d\,\overline u}{du}
\frac{d\,\overline v}{dv}.
\label{2.12}
\end{equation}
We will use such a change of variables later on.
\medskip

Sometimes one may like to ask for the dynamics in a region of
minisuperspace where $a_1$ is large whereas $a_2$ is small.
In such a case, the curvature contribution stemming from
${\bf S}^3$ may be neglected, and $W$ is replaced by
\begin{equation}
W_{\rm approx} = -u^{3/2} + 2\, u^{5/4} v^{1/2}
\label{2.13}
\end{equation}
in (\ref{2.9}) and (\ref{constr}-\ref{2.23/2}).
One would have obtained this
as the correct potential, had one started with a
${\bf T}^3 \times {\bf S}^6$ model from the outset. In this case, one
would replace $d\sigma_3^2$ in (\ref{2.1}) by the metric on the
flat three-torus. Letting the coordinate on the unit torus
range from 0 to $(2\pi^2)^{1/3}$, all the previous formulae
remain valid if the ${\bf S}^3$ curvature contributions are omitted,
including (\ref{2.6}) and (\ref{2.8}). One may even interpret
the model based on $W_{\rm approx}$ as having
${\bf R}^3 \times {\bf S}^6$ as spatial sections, but in this case a
proper integration over ${\cal M}_{9}$ as in (\ref{2.2}) is of
course not possible. One would then ignore (\ref{2.6}) and (\ref{2.8}),
but otherwise arrive at the same equations of motion by directly
using Einstein's field equations. However this case is interpreted,
the equations of motion will leave the absolute scale of $a_1$
free to an arbitrary rescaling
$a_1 \rightarrow c\, a_1$ (i.e. $u^5 v^{-2}
\rightarrow c^8\, u^5 v^{-2}$),
while $a_2 \rightarrow a_2$ (i.e. $v^2 u^{-1}
\rightarrow v^2 u^{-1}$).
\medskip

Summarizing, let us state that we will work in the framework
defined by the structure of the action (\ref{2.9}) and the
equations of motion (\ref{constr}-\ref{2.23/2}),
possibly redefined by a
transformation of the type (\ref{2.12}). The potential (\ref{2.10})
represents the full ${\bf S}^3 \times {\bf S}^6$ model,
whereas (\ref{2.13}) may be viewed either as an approximation
in regions of large external space, or as exact version of the
${\bf T}^3 \times {\bf S}^6$ or ${\bf R}^3 \times {\bf S}^6$
model.

\section{Compactification by signature change}
\setcounter{equation}{0}

One key feature of the model described in the previous Section
is the interplay between the ''causal structure'' of the
Wheeler-DeWitt metric and the sign of the potential $W$. The
set of pairs $(u,v)$ for which $W=0$ is a curve lying entirely
in the interior of minisuperspace. Its explicit form
is exhibited by solving for $v$ as a function of $u$, thus
giving rise to two ''branches'', an ''upper'' one,
\begin{equation}
v = -u^{3/2} +
2 u^2 \Big(\sqrt{u} + \sqrt{u-1}\,\Big) \leadsto 4 u^{5/2}
\label{3.1}
\end{equation}
and a ''lower'' one
\begin{equation}
v = -u^{3/2} + 2 u^2
\Big(\sqrt{u} - \sqrt{u-1}\,\Big) \leadsto \frac{1}{4} u^{1/2},
\label{3.2}
\end{equation}
both branches being connected smoothly at $u=v=1$. The asymptotic forms
given above apply for large $u$, i.e. far away from the origin,
and the second one is at the same time the exact solution
of $W_{\rm approx}=0$.
The whole curve may be parametrized as $u(\lambda)=\cosh^2(\lambda),
v(\lambda)=\cosh^3(\lambda) \exp(2\lambda)$, $\lambda$ taking all
real values. Asymptotically, along the upper branch
$a_1 \rightarrow \ell_{\Lambda} \sqrt{3}$
(while $a_2 \rightarrow \infty$),
along the lower one $a_2 \rightarrow \ell_{\Lambda} \sqrt{15}$
(while $a_1 \rightarrow \infty$). The dashed curve in Fig.1
shows the location of this curve. The domain near the origin
($u$,$v$ small) has $W<0$.
\medskip

The curve $W=0$ divides minisuperspace more or less naturally
into a ''Lorentzian'' region $W>0$ and a ''Euclidean'' region
$W<0$. This notation is motivated by the fact that Lorentzian
trajectories may not emerge from the ''regular zero-geometry''
$u=v=0$, and Euclidean trajectories inside the $W>0$ domain
must necessarily hit the zero potential curve and thus cannot
evolve towards nine-geometries with arbitrarily large volume.
The behaviour of the two types of trajectories
is related to their role in quantum cosmology
\cite{qucos2}. There, one usually
constructs a path integral in the region $W<0$ around Euclidean
trajectories which describe regular ten-geometries. Near the
spacelike part of $W=0$, a family of Lorentzian trajectories that
are supposed to
represent the classical evolution of the universe is
defined by WKB-techniques.
\medskip

Euclidean trajectories corresponding
to regular ten-geometries (by virtue of (\ref{2.1})) behave like
$v \sim c_1 u^{5/2}$ (then $a_1(0)={\rm finite}$) or
$v \sim c_2 u^{1/2}$ (then $a_2(0)={\rm finite}$)
near the origin. There are two preferred solutions (the instantons)
that display high symmetry: One of these is given by the piece
of the curve $v = (9/16) u^{5/2}$
inside the domain $W<0$,
has $a_1 = \ell_{\Lambda} \sqrt{8}$
and corresponds to the ten-geometry ${\bf S}^3 \times {\bf S}^7$.
The other one is given by $v = (4/9) u^{1/2}$
inside $W<0$,
has $a_2 = \ell_{\Lambda} \sqrt{20}$ and describes
${\bf S}^4 \times {\bf S}^6$.
Both solutions have turning points (i.e. $\dot{u}=\dot{v}=0$ in the
gauge $N=1$) at $W=0$.
\medskip

%
%
A generic Euclidean trajectory starting from the origin is confined
to satisfy $du dv>0$ in the Euclidean region. This (and
actually the stronger property $du>0, dv>0$) follows from the
constraint equation (\ref{constr}) with the lower sign, together
with the fact that $W<0$. The trajectory will eventually hit the
curve $W=0$, thereby having either horizontal ($dv=0$) or
vertical ($du=0$) tangent. When evolved further into the
Lorentzian region $W>0$ (but still as a Eulidean trajectory), the
sign change of $W$ will enforce $du dv<0$, and hence drive the
evolution towards the curve $W=0$ again. Re-entering the Euclidean
region, the trajectory evolves with $du dv >0$ (actually $du<0, dv<0$),
until it hits one of the axes and thus describes a Kasner-type
final singularity (i.e. $a_1 \rightarrow 0, a_2 \rightarrow \infty$
or vice versa). In the very special case of the instantons,
the two intersection points between the trajectory and the curve
$W=0$ coincide and form one single turning point, the evolution
leading back to the origin (according to the regularity of the
ten-geometry described by these two instanton solutions).
\medskip

In contrast, Lorentzian trajectories can (due to the constraint
equation (\ref{constr}) with the upper sign) never emerge from
the origin. Hence we restrict our attention to the Lorentzian
trajectories starting at the curve $W=0$ (with -- generically --
$du dv=0$, hence horizontal or vertical tangent). We encounter
two classes of behaviour: Trajectories that evolve towards
arbitrarily large geometries (i.e. values of the spatial
volume) and trajectories than don't (but instead ''recollapse''
towards one of the axes).
\medskip

The first class (i.e. those
Lorentzian trajectories that may represent a reasonable
classical behavior of the universe -- regardless of the space-time
dimension, for the moment) fall into two sub-classes: In the generic
case, one gets exponential inflation in both scale factors $a_1$ and
$a_2$. However, there are two isolated solutions for which one of
the two scale factors is constant. They match the two instanton
solutions at $W=0$ (where they have turning points just as these),
and lie on the $W>0$ pieces of the curves
$v = (9/16) u^{5/2}$ and $v = (4/9) u^{1/2}$,
respectively. Note however that these solutions, although
describing compactification of either scale factor, are not stable
against small perturbations.
\medskip

The second class of Lorentzian
trajectories (i.e. those who do not evolve towards large geometries)
is usually not considered as realistic: They enter
the Lorentzian region, but re-approach the curve $W=0$, hence
leave the Lorentzian region and eventually recollapse towards one
of the axes ($u=0$ or $v=0$) in a Kasner-type singularity
(i.e. $a_1 \rightarrow 0, a_2 \rightarrow \infty$ or
vice versa).
%
%
One may say that a universe described by such a
recollapsing trajectory is ''created'' with too small an
amount of kinetic energy as to become arbitrarily large.
However, this is only a valid statement in the framework
of a model in which one does not admit a change of the
metric signature, once the universe is Lorentzian. In the
approach we are advocating here, those pieces of the
recollapsing Lorentzian trajectories which lie inside
the Lorentzian region, will play a dominant role.
\medskip

In quantum cosmology,
the interplay between Euclidean trajectories (representing in some
sense a full quantum or tunneling state of the universe)
and Lorentzian trajectories (representing a semiclassical state)
is quite implicit, and there is no individual one-to-one
matching. One may, as a different point of view,
regard the transition from one type
of evolution to the other as a {\it classical phenomenon}, thereby
matching a particular Euclidean trajectory to a particular
Lorentzian one. The ten-geometry resulting from such a process
will then, by virtue of (\ref{2.1}), undergo a {\it classical
signature change}. There has been an extensive discussion in the
recent literature whether such a transition is physically
reasonable \cite{EllisSumeruketal}--\cite{sch1},\cite{sch2}--\cite{sch3}.
In the context of cosmology, it is usually conceived to have
happened in the early universe, but its relation to a
''quantum signature change'' as described by the Euclidean
path integral formulation of quantum cosmology is not quite
clear.
\medskip

Taking this possibility serious, we are led to the question under
which conditions trajectories may change their signature type
without rendering the resulting ten-metric too singular.
The answer is to some extent a matter of taste, and we will
require continuity in $\dot{u}/N$ and
$\dot{v}/N$. Then the constraint equation (\ref{constr}) implies that
a classical change of signature can only happen at
points $(u,v)$ for which $W=0$.
\medskip

Let us note for completeness that this
is not the only reasonable choice:
In the most restrictive
versions of classical signature change one requires that the
extrinsic curvature vanishes at the matching nine-surface
\cite{sch2}--\cite{sch3}.
(In Ref. \cite{FE4}, this scenario has been called {\it strong}
signature change).
The extrinsic curvature is in our model essentially given by
the traces over the two factor spheres
\begin{eqnarray}
K_1 &=& \frac{1}{2 {\cal N}}\, h^{ij}_{(1)} \dot{h}_{ij}^{(1)}=
    \frac{3}{\cal N}\,
    \frac{\dot{a}_1}{a_1}=
    \frac{\sqrt{3 \Lambda}}{8 N}
    \,\frac{v^{1/4}}{u^{1/8}} \,
    \Big( 5\,\frac{\dot{u}}{u} - 2\,\frac{\dot{v}}{v} \Big),
\label{3.4}\\
K_2 &=& \frac{1}{2 {\cal N}}\, h^{ij}_{(2)} \dot{h}_{ij}^{(2)}=
    \frac{6}{\cal N}\,
    \frac{\dot{a}_2}{a_2}=
    \frac{\sqrt{3 \Lambda}}{4 N}
    \,\frac{v^{1/4}}{u^{1/8}} \,
    \Big( 2\,\frac{\dot{v}}{v} - \,\frac{\dot{u}}{u} \Big).
\label{3.9/2}\\
\nonumber
\end{eqnarray}
Using the gauge $N=1$ (or any gauge which ensures
$N \neq 0$ when the trajectory approaches $W=0$), the condition $K_i=0$
together with the constraint (\ref{constr}) at $W=0$
implies $\dot{u}=\dot{v}=0$, hence the
existence of turning points for both partners. This is however a
rare event: Among the Euclidean trajectories describing
regular ten-geometries only the two instanton solutions
mentioned above are run into a turning point. Matching these
to the corresponding Lorentzian solutions provides a saddle point
approximation to the Euclidean path integral in quantum cosmology
\cite{qucos2}
and leads (semiclassically) to an (unstable)
compactification of ${\bf S}^3$ or (preferably) ${\bf S}^6$.
\medskip

There are, however, approaches that are less inspired
by quantum cosmology and allow for weaker junction
conditions
\cite{EllisSumeruketal}--\cite{sch1}.
In the spirit of these, we only require
continuity of the extrinsic
curvature.
This is actually identical to assuming continuity of
$\dot{u}/N$ and $\dot{v}/N$.
(In Ref. \cite{FE4}, the scenario based on these
junction conditions has been called
{\it weak} signature change).
To be more explicit, consider,
in the gauge $N=1$, a trajectory of either type approaching
the zero potential curve. Leaving aside the rare possibility of
turning points, one of the two quantities $\dot{u}$, $\dot{v}$
must become zero. If, e.g. $\dot{v}=0$ (horizontal tangent),
this trajectory may be matched to one of the opposite type having
$\dot{v}=0$ as well, such that $\dot{u}$ of the resulting
''mixed'' trajectory is continuous.
%
%
There has recently been some controversy about this
approach in the literature. We will justify its use after
having written down the action (below equation (\ref{3.13/2})).
\medskip

Let us add here a remark in order to avoid confusion. In
Ref. \cite{EllisSumeruketal}, which is one of the most
important papers on the ''weaker'' approach to classical
signature change, a situation similar to ours
(pure gravity with a cosmological constant) is considered
as an example, but within a FRW model containing only a single
scale factor $R$. In this case, the constraint equation
is of the structure $\dot{R}^2/N^2 = \pm W(R)$ instead of
(\ref{constr}). Hence, $W=0$ implies $\dot{R}/N = 0$,
which means vanishing extrinsic curvature. This statement
does {\it not} carry over to the multidimensional case
we are considering. As a consequence, Ellis {\it et al}
precisely recover the classical metric that corresponds
to the Hartle-Hawking no-boundary prescription
\cite{HartleHawking}, i.e. half of ${\bf S}^4$ (the instanton)
matched to half of the deSitter hyperboloid (representing
the classical evolution), whereas we are free to admit a variety
of signature change configurations, the instanton trajectories
providing just a very special isolated case.
\medskip

As a second assumption, we not only {\it admit} the possibility of
signature change but require that it {\it will} happen whenever a
trajectory approaches the curve $W=0$. In terms of a single
expression, such mixed trajectories are produced by the action
\begin{equation}
S = \frac{1}{\Lambda_{\rm{eff}}^4}
   \int dt \left( -\, \frac{\dot u \dot v}{N}
                 -  N |W(u,v)| \right).
\label{3.5}
\end{equation}
Denoting $s \equiv {\rm sgn}(W)$, hence $|W| \equiv s W$, the equations
of motion are given by
\begin{eqnarray}
\frac{\dot{u}\dot{v}}{N^2} & = & s\, W,\label{constrs}\\
\frac{1}{N}\frac{d}{dt}\bigg(\frac{\dot{u}}{N}\bigg) & =
            & s\, \partial_v W,\label{3.6}\\
\frac{1}{N}\frac{d}{dt}\bigg(\frac{\dot{v}}{N}\bigg) & =
            & s\, \partial_u W.\label{3.13/2}\\
\nonumber
\end{eqnarray}
This form is suitable for numerical methods as well. One may
of course replace $s \partial W$ by $\partial |W|$.
\medskip

%
%
Here, a further remark on the junction conditions is in place.
Recently, it was claimed that the original relaxation of
jump conditions as advocated in Ref. \cite{EllisSumeruketal}
does not correctly take into account the distributional nature of
Einstein's field equations (see e.g. Ref. \cite{sch4} and the first of
Ref. \cite{sch3}).
Hence, in the strict sense, Einstein's equations are not satisfied
at the hypersurfaces of signature change. Thereby we denote by
''Einstein's equations'' those derived from the action (\ref{2.2})
with the {\it upper} sign for {\it both} signature types.
In the minisuperspace
model under consideration, this would give rise to a
Lagrangian proportional to $ - s\, \dot{u}\dot{v} / N - N W$,
instead of (\ref{3.5}).
However, there is another possibility to obtain a model for
signature change, namely
by inserting an additional minus sign for the
Euclidean case. This is indicated by the double sign in (\ref{2.2})
and leads to (\ref{3.5}),
hence to a Lagrangian proportional to
$ -  \dot{u}\dot{v} / N - s N W$.
The difference between these two aproaches at the level of the
field equations consists essentially of $\dot{s}$-terms,
hence $\delta$-distributions on the hypersurface of signature change.
As is clear from (\ref{constrs}-\ref{3.13/2}),
such terms do not arise in the
second approach. Thus we end up with a model
described by the action (\ref{3.5}), implying
the junction conditions delevoped and
exploited in Refs.\cite{EllisSumeruketal}--\cite{sch1}.
\medskip

This line of reasoning is not restricted to the FRW model
we consider here, but follows a general pattern.
In Ref. \cite{FE4}, the junction conditions demanding continuity
of the extrinsic curvature were denoted as the {\it weak} ones
and have been studied in the context of a
classification of possible covariant action integrals for
signature change.
The general justification of such a model arises from the fact
that the Einstein-Hilbert action for the Euclidean part of
the space-''time'' manifold may in general be assumed
to have either the {\it same} or the {\it opposite} sign
as compared to the Lorentzian Einstein-Hilbert action,
\begin{equation}
S \sim \mp \int_{{\cal M}_{\rm eucl}} d^n x\,\sqrt{|g|}\, R
     - \int_{{\cal M}_{\rm lor}} d^n x\,\sqrt{|g|}\, R.
\label{xyz}
\end{equation}
Both of these two models are thus defined by generally
covariant actions, and since we do not know the
''correct'' sign in front of the Euclidean Ricci scalar,
they are {\it \`a priorily} of equal right. (Some subtleties
like how to integrate across the hypersurface of signature
change, and the possible inclusion of boundary integrals
are discussed in detail in Ref. \cite{FE4}).
The first
possibility (equal signs) leads to {\it strong} signature change,
the second one (opposite signs) leads to {\it weak} signature
change, and it is this second one which we choose as the underlying
theory.
\medskip

We will now show that the {\it weak} signature change approach
as formulated above leads to a new type of compactification
mechanism. Begin with a Euclidean trajectory that is supposed
to describe the ''birth'' of the very early universe. (It need
not represent a regular ten-geometry in this context, but it will
at least be reasonable to let it start from the origin $u=v=0$).
The trajectory eventually approaches the curve $W=0$, where it
is matched to a Lorentzian solution according to our prescription.
As mentioned above, it may happen that this trajectory does
not escape towards arbitrarily large nine-space volumes but
re-approaches $W=0$ and thus re-enters the Euclidean domain.
Instead of recollapsing (as is usually assumed), the
trajectory becomes Euclidean there by signature change,
and will in turn re-appear in
the Lorentzian region, where it becomes Lorentzian
again. This process may continue indefinitely.
%
%
In summary, it consists of assuming the trajectory to be
of Lorentzian type in the Lorentzian region, and of
Euclidean type in the Euclidean region.
In the generic case, the resulting trajectory ''oscillates'' in
signature type and remains always near one of the two branches
(\ref{3.1}--\ref{3.2}) of $W=0$. The corresponding ten-metric
(consisting of pieces (\ref{2.1})) undergoes an infinite sequence
of signature changes.
\medskip

We recall that the two branches of the zero potential curve
correspond asymptotically to constant values of one of the
scale factors. Approximately half of all mixed trajectories
oscillate around the lower branch, and thus undergo a ''time''
evolution $a_1(t) \rightarrow \infty$,
$a_2(t) \rightarrow \ell_{\Lambda} \sqrt{15}$.
For these solutions, ${\bf S}^6$ compactifies to the scale set
by the cosmological constant $\ell_{\Lambda} \equiv \Lambda^{-1/2}$.
Alternatively (though physically less desirable), ${\bf S}^3$
may compactify as well.
\medskip

Fig.1 presents an example of a mixed trajectory describing
compactification of ${\bf S}^6$. It emerges from the origin
as $u^{-1/2} v \rightarrow 0.5$
(corresponding to a regular ten-geometry) and
has been evolved numerically in the gauge $N=1$. The dashed
curve is $W=0$. Fig.2 shows the dependence of the two
scale factors $a_1$, $a_2$ on the ''time'' parameter $t$
(the unit on the vertical axis is $\ell_{\Lambda}$).
In Fig.3, the graph of the function $a_2(t)$ is displayed
using a finer resolution.
This numerical solution suggests that $a_1(t)$ oscillates around
some linearly growing average, whereas $a_2(t)$ performs
damped oscillations around its limiting value
$\ell_{\Lambda} \sqrt{15}$.
\medskip

The stability of this compactification scheme is evident,
since small perturbations in the scale factors do not
alter the qualitative behaviour of mixed trajectories. This
is due to the fact that the configurations we are
talking about do not provide isolated
solutions. In other words, given a reasonable measure
on the set of all possible initial (Euclidean) trajectories
that emerge from the origin, compactification of either
scale factor will occur with finite probability.
\medskip

However nice these pictures look, the most important questions
remain to be answered: What is the (effective, i.e.
physically measurable) metric, and why do we experience it as a
Lorentzian rather than a Euclidean one? What is the evolution
of the large scale factor in terms of a physical time
coordinate? These questions can be answered indeed, and the next
Section is devoted to them.

\section{Effective space-time metric}
\setcounter{equation}{0}

The metric described by a mixed oscillating trajectory may be
expressed in terms of the ''proper time'' gauge ${\cal N}=1$
(the corresponding coordinate being denoted $\tau$). It
consists of pieces (\ref{2.1}) with alternating signs. Let
us write such a metric as
\begin{equation}
ds^2 = - \,s(\tau) d\tau^2 + a_1(\tau)^2 d\sigma_3^2 +
       a_2(\tau)^2 d\sigma_6^2,
\label{4.1}
\end{equation}
where $s \equiv {\rm sgn}(W) = \pm 1$. (The coordinate $\tau$
is identical to what is called $\sigma$ in
Ref. \cite{EllisSumeruketal}).
Let the signature change occur
at values $\tau_j$ ($j=1,2,3...$) of the time parameter. We will
adopt the convention that the Lorentzian intervals are given
by $\tau_{j-1}<\tau<\tau_j$ for {\it even} integers $j$. Hence,
$\Delta\tau_j = \tau_j - \tau_{j-1}$ is a Lorentzian type parameter
time interval if $j$ is {\it even}, and a Euclidean type interval
if $j$ is {\it odd}.
\medskip

Since the size of these intervals is of the order $\ell_{\Lambda}$
at the beginning (one usually assumes $\ell_{\Lambda} \approx
{\rm compactification \, radius} > {\rm or} \approx \ell_P$),
we expect some average over (\ref{4.1}) to yield
the effective physical space-time metric. The most natural
measure for such an average procedure is provided by the
proper ''time'' coordinate $\tau$ as used in (\ref{4.1}) and
its intervals $\Delta\tau_j$, although what follows is quite
robust against the use of a different time coordinate. Hence,
we assume the effective metric for large $\tau$ to be
\begin{equation}
ds^2_{\rm eff}=g_{00}^{\rm eff}(\tau)d\tau^2 +
a_1(\tau)^2 d\sigma_3^2 + a_2(\tau)^2 d\sigma_6^2
\label{4.3/2}
\end{equation}
with $a_2^{\rm eff}(\tau) = \ell_{\Lambda} \sqrt{15} \equiv
(15/\Lambda)^{1/2}$ and
\begin{equation}
g_{00}^{\rm eff}(\tau) = -\, \frac{\sum (-)^j\Delta\tau_j}
                           {\sum \Delta\tau_j},
\label{4.2}
\end{equation}
where the sum is over a large number of intervals located
near $\tau$. Due to the qualitative behaviour of the mixed trajectory
and the scale factors as displayed in Fig.1 and Fig.2,
one would expect the Lorentzian and the Euclidean contributions
to $g_{00}^{\rm eff}$ to be of equal size and thus to cancel,
giving $g_{00}^{\rm eff}=0$. This is however only true in the
limit $\tau \rightarrow \infty$, and we will show in the following
that the actual behaviour of the metric is
$g_{00}^{\rm eff} \sim \tau^{-2}$ for large $\tau$.
\medskip

In order to apply analytic methods to some relevant order, we
consider a mixed oscillating trajectory that has already
evolved along the lower branch of $W=0$ into the region with
large $a_1$. (A similar
computation is of course possible for the upper branch, in
which case $a_1$ compactifies). As mentioned in Sect.2,
the dynamics is well described by $W_{\rm approx}$ from
(\ref{2.13}). The structure of this approximation is most
easily exhibited by performing a change of variables of
the type (\ref{2.12}). Let us call the new variables
($x,y,\widetilde{N}$) and set
\begin{equation}
u=x^{2/3}, \qquad v=y^{1/3}.
\label{4.3}
\end{equation}
Furthermore, we define
\begin{equation}
\frac{y}{x} = \bigg(\frac{v}{u^{1/2}}\bigg)^3 \equiv \zeta
    \equiv z^6.
\label{4.4}
\end{equation}
The potential arising is given by
\begin{equation}
\widetilde{W} = W_{\rm approx}\, \frac{du}{dx}\frac{dv}{dy} =
   \frac{2}{9} \Big( -\zeta^{-2/3} + 2\, \zeta^{-1/2} \Big)
   \equiv \widetilde{W}(\zeta).
\label{4.5}
\end{equation}
The new lapse variable $\widetilde{N}$ relates to the previous ones
by
\begin{equation}
{\cal N}=\frac{3\sqrt{10}}{\Lambda a_2} \,N =
\sqrt{\frac{3}{\Lambda}}\, \zeta^{-1/12}\, N =
\frac{2}{9} \sqrt{\frac{3}{\Lambda}}\, x^{-1}\, \zeta^{-3/4}\,
       \widetilde{N},
\label{4.6}
\end{equation}
and for convenience we note the transformation formulae
\begin{eqnarray}
a_1&=&\sqrt{\frac{6}{\Lambda}}\, u^{5/8}\, v^{-1/4} =
    \sqrt{\frac{6}{\Lambda}}\, x^{1/3}\, \zeta^{-1/12} =
    \sqrt{\frac{6}{\Lambda}}\, x^{1/3}\, z^{-1/2},
            \label{4.7}\\
a_2&=&\sqrt{\frac{30}{\Lambda}}\, u^{-1/8}\, v^{1/4} =
    \sqrt{\frac{30}{\Lambda}}\, \zeta^{1/12} =
    \sqrt{\frac{30}{\Lambda}}\, z^{1/2}.
            \label{4.15/2}\\
\nonumber
\end{eqnarray}
The sign of $\widetilde{W}$ translates into
$s = {\rm sgn}(\zeta-\zeta_0) =
{\rm sgn}(z-z_0)$, with $\zeta_0 = 1/64$, $z_0=1/2$ representing
the zero potential curve as well as the limiting value of $a_2$.
\medskip

The action and the equations of motion are now of the type
(\ref{3.5}-\ref{3.13/2})
with $u \rightarrow x$, $v \rightarrow y$, and
$W \rightarrow \widetilde{W}$ replaced.
The fact that $\widetilde{W}$
depends only on the ratio $y/x$ (and is thus homogenous of
degree zero) corresponds to the fact that the absolute
scale of $a_1$ is free to rescalings (cf. the remarks at the
end of Sect.2). Due to this symmetry, the equations of motion
simplify if we choose the gauge
$\widetilde{N} = x \zeta^{3/4}$.
This corresponds to ${\cal N} = (2/9)(3/\Lambda)^{1/2}$, which
implies
\begin{equation}
\tau = \frac{2}{9} \sqrt{\frac{3}{\Lambda}}\, t.
\label{4.8}
\end{equation}
For convenience, we will work with $t$ instead of $\tau$
($t_j$ and $\Delta t_j$ being defined in an obvious way).
\medskip

It turns out that, when written in terms of $x$ and $z$ as
independent variables, the equations of motion contain $\dot{x}$
but not $x$. Setting
\begin{equation}
\sigma = \frac{\dot{x}}{x},
\label{4.9}
\end{equation}
and performing some algebra, they take the form
\begin{eqnarray}
\dot{z} & = & -\,\frac{1}{6}\, z \sigma + \frac{s}{27\sigma}
                     \Big( 2 z-1 \Big),
               \label{4.10}\\
\dot{\sigma} & = & -\,\frac{3}{4}\, \sigma^2 + \frac{s}{54}
                    \Big( 6 - \frac{1}{z} \Big),
               \label{4.11}\\
\ddot{z} & = & - \,\frac{\dot{z}^2}{2 z} - \sigma \dot{z}
               + \frac{4 s}{81} \Big( \frac{3 z}{2} -1 \Big).
              \label{4.12}\\
\nonumber
\end{eqnarray}
The structure of the equations has now changed:
(\ref{4.10},\ref{4.11})
define a two-dimensional dynamical system, whereas (\ref{4.12})
is a consequence thereof and may be omitted (In numerical
computations,
the use of equations (\ref{4.11},\ref{4.12}) which are not
singular at $\sigma=0$ may turn out to be more appropriate.
In this case, (\ref{4.10})
gives the initial value for $\dot{z}$ if those of $z$ and
$\sigma$ are prescribed). The appearance of such
reductions in simple cosmological models is well known
\cite{dyn}.
\medskip

Since (\ref{4.10}-\ref{4.12}) describe escaping trajectories as well
as oscillating ones, one has to prescribe appropriate
initial values at some time $t_{\rm ini}$. In Fig.4,
an example (with $z(0)=0.6,\, \sigma(0)=0.31$) is displayed.
The dashed line represents $\sigma(t)$, the solid one shows
(using a magnification factor of 5 in order
to keep common units on the vertical axis)
the function $ 5\,(z(t)-1/2)$.
During the Lorentzian periods, $\sigma$ increases, during the
Euclidean periods, it decreases to zero. Its maxima
decrease and converge to zero for $t \rightarrow \infty$.
$z$ performs damped oscillations around its limiting value $z_0 = 1/2$.
Equation (\ref{4.11}) tells us that $\sigma(t)$ tends
to a zigzag curve with slope $\pm 2/27$.
\medskip

Our main concern is the estimation of the ''time'' intervals
$\Delta t_j$ and the long-time behaviour of $a_1(t)$. Let
us solve the equations of motion near some value $t = t_{\rm ini}$
that is large enough for our approximation to hold.
The most convenient choice is $t_{\rm ini} \equiv t_j$
for some {\it odd} integer $j$. In other words, we place the
initial time at the end of a Euclidean interval (at the beginning
of a Lorentzian one). The interval $I_j$, defined by
$t_{j-1}<t<t_{j+1}$, is
understood as a ''pair'' of two periods of the type
(eucl,lor), and it is followed by
another pair $t_{j+1}<t<t_{j+3}$ of the same type, and so on.
A power series ansatz reveals that $q \equiv \dot{z}(t_j)$
is the only free parameter ($z=1/2$ and $\sigma=0$ there).
Within the interval $I_j$, we may use $s \equiv {\rm sgn}(t-t_j)$.
Denoting $ \widetilde{t} \equiv t-t_j$,
the solution reads, to within the order that it necessary for the
applications we have in mind,
\begin{eqnarray}
z(t) & = & \frac{1}{2} + q\, \widetilde{t} + A\, \widetilde{t}^{\,2}
        + B\, \widetilde{t}^{\,3} + D\, \widetilde{t}^{\,4}
        + ...\,\, ,
                       \label{4.13}\\
\sigma(t) & = & \frac{2 s}{27}\, \widetilde{t}
            + \frac{s q}{27}\, \widetilde{t}^{\,2}
      + E\, \widetilde{t}^{\,3} + F\, \widetilde{t}^{\,4}
      + ... \,\, ,
                       \label{4.14}\\
\nonumber
\end{eqnarray}
where
\begin{eqnarray}
A &=& -\, \frac{1}{162} ( s + 81\, q^2),
                       \nonumber\\
B &=& \frac{1}{243} ( s q + 162\, q^3),
                       \nonumber\\
D &=& \frac{1}{78732} (2 - 729\, s q^2 - 91854\, q^4),
                       \nonumber\\
E &=& - \,\frac{5}{6561} (2 + 81\, s q^2),
                       \nonumber\\
F &=& -\, \frac{1}{26244} ( 13\, q - 3240\, s q^3).
                       \nonumber\\
\nonumber
\end{eqnarray}
Recall that this solutions applies only in the interval $I_j$.
Clearly, in the succeeding interval $I_{j+2}$, which is
again of the type (eucl,lor), a similar solution
applies (with $q \rightarrow q' \equiv \dot{z}(t_{j+2})$ and
$\widetilde{t} \rightarrow t-t_{j+2}$ replaced).
\medskip

The size of the two periods that make up $I_j$ may now be computed
as the zeros of $z(t)-1/2$ as given by (\ref{4.13}).
For the sake of notational ease, we will use the
equality sign to indicate the first relevant order(s) of
various quantities. It turns out that
\begin{eqnarray}
\Delta t_{\rm eucl} & = & \Delta t_j =
             162\, q - 21870\, q^3,
                       \label{4.15}\\
\Delta t_{\rm lor} & = & \Delta t_{j+1} =
             162\, q + 21870\, q^3.
                       \label{4.16}\\
\nonumber
\end{eqnarray}
This shows a small amount of asymmetry between the Lorentzian and
the Euclidean interval sizes that will be important below.
\medskip

The period sizes inside the interval $I_{j+2}$ shall be
denoted as $(\Delta t_{\rm eucl}',\Delta t_{\rm lor}')$. Using
(\ref{4.13}), one finds
\begin{equation}
\dot{z}(t_{j\pm 1}) = -q \pm 324\, q^3.
\label{4.17}
\end{equation}
%
%
This provides a characterization of the Taylor expansions we used:
The series are truncated at an order such as to reproduce
(\ref{4.15})--(\ref{4.17}) correctly, where it is  understood that
$O(q^4)$-contributions are neglected.
Thus, to the order considered,
\begin{equation}
q'= q - 648\, q^3
\label{4.18}
\end{equation}
plays exactly the same role for the interval $I_{j+2}$
as $q$ does for the interval $I_j$
(recall $q' \equiv \dot{z}(t_{j+2})$). Repeating
(\ref{4.15},\ref{4.16})
for this new interval, we obtain the period sizes in terms
of $q$:
\begin{eqnarray}
\Delta t_{\rm eucl}' & = & \Delta t_{j+2} =
             162\, q - 126846 \,q^3,
                       \label{4.19}\\
\Delta t_{\rm lor}' & = & \Delta t_{j+3} =
             162\, q - 83106 \,q^3.
                       \label{4.20}\\
\nonumber
\end{eqnarray}
\medskip

These results provide enough information to answer all the questions
posed so far. Denoting $q_j \equiv \dot{z}(t_j)$, equation
(\ref{4.18}) tells us
\begin{equation}
q_{j+2}-q_j = - 648\, q_j^3.
\label{4.21}
\end{equation}
For large j, we can rewrite this as the differential equation
$dq/dj = -324 q^3$ and obtain to leading order
\begin{equation}
q_j = \frac{1}{18 \sqrt{2 j}}.
\label{4.22}
\end{equation}
This implies that the period sizes are
\begin{equation}
\Delta t_j = \frac{9}{\sqrt{2 j}}.
\label{4.23}
\end{equation}
We will add a next-order correction to this expression later on.
These numbers can be summed over to provide (to leading order)
the value $t_j$ corresponding to the $j$-th signature change,
\begin{equation}
t_j = \frac{9}{\sqrt{2}} \sum_{k=1}^j k^{-1/2}
    = 9 \sqrt{2 j}.
\label{4.24}
\end{equation}
Note that the deviations from (\ref{4.23}) that stem from
the early evolution where the approximation
$W \approx W_{\rm approx}$ is not valid, contribute at most
an additive constant to $t_j$, and thus are irrelevant.
\medskip

An immediate consequence of this is (using the zigzag limit
of $\sigma(t)$ inside each interval $I_j$)
\begin{equation}
\frac{1}{3} \int_0^{t_j} dt' \sigma(t') =
\frac{1}{81} \sum_{k=1}^j \Delta t_k^2 =
\frac{1}{2} \sum_{k=1}^j \frac{1}{k} =
\frac{1}{2} \ln j =
\frac{1}{2}\ln t^2 =
\ln t.
\label{4.25}
\end{equation}
Hence, with (\ref{4.9}), we obtain the long-time behaviour
$x(t) \sim t^3$. Inserting this into
(\ref{4.7}) and letting $z \rightarrow 1/2$,
we find for large $t$
\begin{equation}
a_1(t) = \widetilde{C} t,
\label{4.26}
\end{equation}
where $\widetilde{C}$ is a constant. This has been already anticipated in the
previous Section.
\medskip

The most important result, however, is contained in the equations
(\ref{4.15},\ref{4.16}) and
(\ref{4.19},\ref{4.20}). They provide a subtle
pattern of small perturbations to (\ref{4.23}). Looking at the
different signs showing up in these four expressions, one may recast
the general behavior into the form
\begin{equation}
\Delta t_j = \frac{9}{\sqrt{2 j}} +
           (-)^j \frac{33}{8 \sqrt{2}\, j^{3/2}}.
\label{4.27}
\end{equation}
This equation, along with (\ref{4.22}) and (\ref{4.24}),
is valid for odd as well as for even integers $j$.
It displays a small predominance of the Lorentzian
period sizes ($j$ even) over the Euclidean ones ($j$ odd).
\medskip

As a last application of this machinery, we may compute the
amplitudes of the oscillations. Denoting by $\delta$ the
absolute value of the maximal deviations of a quantity from
its long-time limit during one period, we find to leading
order
\begin{eqnarray}
\delta \sigma  &=&  \frac{2}{27}\, \Delta t
     = \frac{1}{3}\sqrt{\frac{2}{j}} = \frac{6}{t},
                       \label{4.28}\\
\delta z  &=& \frac{1}{6}\, \frac{\delta x}{x} =
            \frac{1}{3}\, \frac{\delta a_1}{a_1} =
             \frac{\delta a_2}{a_2} = \frac{81}{2} q_j^2
               = \frac{1}{16 j} = \frac{81}{8 t^2}.
                       \label{4.29}\\
\nonumber
\end{eqnarray}
This shows the amount of damping that occurs to the various
quantities.
\medskip

Having evaluated the necessary ingredients, we return to the
computation of the effective space-time metric (\ref{4.3/2}),
which amounts to perform some average procedure of the type
(\ref{4.2}) to (\ref{4.27}). Again, we encounter some subtleties.
Naively, one would expect to obtain $g_{00}^{\rm eff}$ as
an average over two neighbouring
time periods. Let us try this for
a pair ($\Delta t_{j-1},\Delta t_j$), with $j$ of either parity.
The prescription (\ref{4.2}) leads to
\begin{equation}
-\frac{ (-)^j \Delta t_j + (-)^{j-1} \Delta t_{j-1} }
      { \Delta t_j + \Delta t_{j-1} } =
\frac{1}{j}  \bigg( \frac{(-)^j}{4} - \frac{11}{24} \bigg).
\label{4.31}
\end{equation}
Hence, an average over a pair of periods of type
(eucl,lor) gives a result different from
an average over (lor,eucl). However, since to first
order the neighbouring $\Delta t$'s are of equal size,
we average over (\ref{4.31}) for two neighbouring
$j$'s (i.e. an even and an odd one). The result is
$ -11/(24 j) = - 297/(4 t^2)$, which is negative and falls
off to zero as $t \rightarrow \infty$. Taking into
account the factor (\ref{4.8}) between $t$ and $\tau$,
the effective metric becomes
\begin{equation}
ds^2_{\rm eff} = -\,\frac{11}{\Lambda} \frac{d\tau^2}{\tau^2}
      + C^2 \tau^2 d\sigma_3^2
      + \frac{15}{\Lambda} d\sigma_6^2.
\label{4.32}
\end{equation}
The constant $C$ may be set to any given value by a trivial
rescaling of $\tau$.
(If the formalism
based on $\widetilde{W}$ that we used in this Section
is interpreted as the exact version of
a ${\bf T}^3 \times {\bf S}^6$ or ${\bf R}^3 \times {\bf S}^6$
model, then the origin of such a freedom is clear.
If, on the other hand, this formalism is considered
as a large-$a_1$ approximation of the
${\bf S}^3 \times {\bf S}^6$ model we started with,
(\ref{4.32}) shows that there is enough loss of information
about the initial trajectory to allow for such
a rescaling as well.)
A further transformation to the effective (physical) proper
time
\begin{equation}
\sqrt{\frac{11}{\Lambda}} \ln\bigg(\frac{\tau}{\tau_0}\bigg) = \eta,
\label{4.33}
\end{equation}
with $\tau_0$ an arbitrary constant, gives
\begin{equation}
ds^2_{\rm eff} = - d\eta^2
      + \frac{1}{\Lambda} \exp \bigg(
         2 \sqrt{\frac{\Lambda}{11}}\eta \bigg) d\sigma_3^2
      + \frac{15}{\Lambda} d\sigma_6^2,
\label{4.34}
\end{equation}
where $\tau_0$ has been used to produce a nice prefactor
of the exponential. (Note that by
$\eta \rightarrow \eta + const$, this prefactor may be
rescaled to any other value).
The effective four-metric is obtained by omitting the
$d\sigma_6^2$-term. Thus, the small predominance of the
Lorentzian over the Euclidean periods as described by
(\ref{4.27}) leads to an effective Lorentzian (i.e.
Pseudo-Riemannian) metric displaying exponential inflation
in the remaining scale factor $a_1$. This is our main result.
As a particularly nice feature typical for inflation we observe
that the metric is unique, hence independent of the initial values
of the trajectory.
\medskip

Let us finally write down the physical
time values at which signature change occurs,
\begin{equation}
\eta_j = \frac{1}{2} \sqrt{\frac{11}{\Lambda}} \ln j,
\label{4.35}
\end{equation}
and the corresponding period sizes
\begin{equation}
\Delta \eta_j = \frac{1}{2 j}\,\sqrt{\frac{11}{\Lambda}} =
          81 \sqrt{\frac{11}{\Lambda}}
          \exp \bigg(-2 \sqrt{\frac{\Lambda}{11}} \eta_j\bigg).
\label{4.36}
\end{equation}
This concludes the presentation of the
signature change model of compactification.
The remaining two Sections are devoted to a discussion
of open questions, speculations and concluding remarks.

\section{Physical problems}
\setcounter{equation}{0}

Although the effective metric (\ref{4.34}) looks appealing,
the way it has been obtained appears somewhat {\it at hoc}.
The two essential steps in our argumentation are on the
one hand the assumption that signature change occurs whenever
it is possible, and on the other hand the prescription
(\ref{4.2}) that amounts to perform some average over the true
metric in order to get the effective one. A possible point
of view about this procedure
is that what we have done is effectively some
approximation or limiting case to a different theory, possibly
connected with some version of quantum gravity. One should
keep in mind that our first assumption (the actual occurence
of signature change) raises the issue of the dynamical interplay
between Lorentzian and Euclidean geometries, whereas the
second one (the average procedure) seems to mimick some
coarse graining (though at a classical level). Both of these
topics are under discussion in recent quantum cosmology as well.
There has even been a proposal of a quantum theory of
what we called
''classical'' signature change \cite{Martin} (that leads to
a modified Wheeler-DeWitt equation). As already mentioned in
Ref. \cite{FE1}, mixed trajectories in superspace might give
rise to semiclassical states in such a theory. The effective
metric (\ref{4.34}) would then probably be a prediction based
on a (non-standard) theory of quantum cosmoloy (at least
for some proper time interval after which the notion of
classical signature oscillations might completely break down
-- see below). Regrettably, we have to leave these very fundamental
questions open.
\medskip

Examining the predictive power of the mechanism we presented,
we observe that the metric (\ref{4.34}) -- although being
unique and nicely describing an inflationary universe --
does not provide a hint how inflation eventually would stop,
and how the evolution of the large scale factor $a_1$
would become a standard
(e.g. radiation or matter dominated) one. Usually, one expects
this sort of problem in a theory containing a positive
cosmological constant. In the standard approaches
\cite{Halliwell3}--\cite{Hawking},
$\Lambda$
mimicks the potential value $V(\phi_0)$ of some initial
scalar field $\phi_0$. An exit out of inflation is provided by
the actual dynamics of $\phi$: As $V(\phi)$ decreases, the
cosmological constant ''decays''. Such a procedure does not seem
to work in the model we are considering. The reason is that
the compactification scale $a_2 \sim \ell_\Lambda \equiv
\Lambda^{-1/2}$ would blow up as $\Lambda \rightarrow 0$,
and the main purpose of the mechanism would be lost. In other
words, in order to retain compactification for all times, the
cosmological constant should be fundamental and not just a
convenient way to mimick matter in the early stages of
the evolution. The problem is here to exit inflation without
switching off $\Lambda$. Let us list some speculations how it
might be overcome.
\medskip

{}From the point of view of any quantum theory of gravity, the
appearance of classical time periods below the Planck scale are
problematic. However, this is precisely what happens in
our model. The time interval between two signature changes
drop down to zero quite rapidly. This is independent on whether
the characteristic scale is expressed in terms of the underlying
''microscopic proper time'' $\tau$
\begin{equation}
\Delta \tau_j = \frac{12\, \ell_\Lambda^{\,2}}{\tau_j}
\label{5.1}
\end{equation}
(which follows from (\ref{4.23},\ref{4.24}) and (\ref{4.8})),
or in terms of the physical proper time $\eta$, as in
(\ref{4.36}). Even if $\ell_\Lambda$ exceeds $\ell_P$ by some
orders of magnitude, the period sizes $\Delta \tau$ become
Planckian at $\tau \approx \ell_\Lambda^{\,2}/\ell_P$, the physical
time intervals $\Delta \eta$ become Planckian at
$\eta \approx \ell_\Lambda \ln(\ell_\Lambda/\ell_P)$. One thus
expects quantum gravity effects to become important at some
stage of the inflationary evolution (\ref{4.34}). These
would certainly modify the model in the sense that the classical
oscillating trajectories are replaced by some prescription
how to compute expectation values. It is not even clear to
what extent WKB techniques would apply, because it is conceivable
(especially if $\ell_\Lambda \approx \ell_P$) that the
universe is permanently in some full (non-WKB) quantum state
as far as $a_2$ is concerned, and only the observable $a_1$
behaves classical. However, one might expect the classical
arguments leading to (\ref{4.26}) to break down
or to become strongly modified so that inflation is
eventually stopped by quantum effects. This touches upon
the fundamental question how physical time is ''created'' from
a quantum state
\cite{AshtekarStachel}.
In other words: the way physical time was
constructed in the previous Section (by averaging with
respect to a ''microscopic'' oscillating-signature type
parameter $\tau$) might break down as the oscillation
periods fall below Planck scale. In addition, a realistic model
would contain matter fields. For example, in the case of a
single scalar field, the total cosmological constant could
be a sum
\begin{equation}
\Lambda_{\rm tot} = \Lambda_{\rm fund} + V(\phi)
\label{5.2}
\end{equation}
of a fundamental and a ''decaying'' part. Only $\Lambda_{\rm fund}$
would survive and guarantee compactification, while the scalar
field would produce density fluctuations and matter particles.
Whether such a modification may be constructed, and whether the
resulting effective four-metric produces an amount of inflation
and structure that is
compatible with observational constraints must be
left for future research.
\medskip

We can add a technical remark that is related to the problem
of constructing semiclassical states around mixed trajectories,
and that has a purely classical counterpart, too. Usually,
one expects a one-dimensional family of trajectories emerging
from a common point in minisuperspace (the origin, say) to allow
for a description in terms of the classical action $S$
(integrated along the trajectories) and a set of
Hamilton-Jacobi equations. In the gauge $N=1$, these are
\begin{equation}
\dot{u} = -\, \partial_v S, \qquad \dot{v}= -\,\partial_u S.
\label{zwischen}
\end{equation}
However, such a formulation is only possible if the different
trajectories do
not intersect each other. In our model, the trajectories
actually {\it do} intersect. One can show that this situation
generates an infinite sequence of functions $S_k$, each
describing only a subset of the family of solutions
by means of (\ref{zwischen}) in a restricted domain of
minisuperspace. In a sloppy manner, one can imagine $S_k$ to be
an evaluation of (\ref{3.5}) along a trajectory having undergone
$k$ oscillations already. This raises the question how the
procedure to assign a semiclassical state
$\psi \sim \exp (i S)$ to a family of classical trajectories
\cite{Halliwell3}
is modified. If, by construction, such a state is based
on the set of partial actions $\{S_k\}$, one might speculate that
some ''damping'' effect alters the effective dynamics
of the large scale factor and hence cures the problem of
eternal inflation.
\medskip

There is, however, a chance to obtain a classical,
non-inflationary long-time
behavior even in a rather standard manner, namely by introducing
additional matter fields and couplings, or by incorpotating
a phenomenological (perfect fluid type) energy momentum tensor
obeying some equation of state. It is conceivable that a
perturbation  of the dynamical system (\ref{4.10}-\ref{4.12})
along these lines alters the amplitudes of the various
quantities during the oscillations. The changes required
are not drastic: Retaining, for example, the relation between
$t$,  $\tau$ and the physical proper time $\eta$, one would
need $a_1 \sim (\ln t)^{\kappa}$ as $t \rightarrow \infty$
instead of (\ref{4.26}). If $0<\kappa<1$,
it follows $a_1 \sim \eta^\kappa$, hence
non-inflationary cosmic expansion.
This may in turn be achieved if the amplitudes of the
oscillations in $\sigma$ behave like
$\delta \sigma = 6 \kappa (t \ln t)^{-1}$ instead of
$\delta \sigma = 6 t^{-1}$, as given by (\ref{4.28}).
In other words, any additional coupling that increases the
damping effects on $\sigma$ works against inflation.
\medskip

There is another possible way out of inflation that even gives
up $\Lambda$ as a fundamental constant: We have implicitly assumed
that the effective four-metric is just the averaged
truncated version of the full ten-metric. This is not
necessarily the case. One may embed the model into a theory
in which the true effective metric is given by $ds_{\rm eff}^2$
(as evaluated in the previous Section)
only up to a conformal factor,
\begin{equation}
ds_{\rm true}^2 = \Omega(\eta)^2 ds_{\rm eff}^2.
\label{5.3}
\end{equation}
One often encounters such a situation in the dimensional
reduction of supergravity and superstring theories
(see e.g. Ref. \cite{CremmerJulia} for an article
explaining the rescaling of various ''metrics'' --
as appearing in supergravity -- in great detail, and
Ref. \cite{GSW} for a bulk of
material on superstrings). The
conformal factor is usually related to some scalar field
(as e.g. a dilaton). Moreover, assume $\Lambda$ to be
an effective energy of some field, typically $V(\phi)$,
such that it decays slowly (''adiabatically''). We indicate
this by writing $\Lambda = \Lambda(\tau)$. Using
$ds_{\rm eff}^2$ in the form (\ref{4.32}), compactification
is retained if $\Omega(\tau) \sim \Lambda(\tau)^{-1/2}$.
The type of three-space expansion is then determined
by the ''constant'' $C$. Although completely arbitrary
in the large-$a_1$ approximation we used, it may to an even
higher approximation be related
to $\Lambda(\tau)$ and the initial values of the
trajectory. As a possible way out of eternal inflation,
one could try to adjust the field contents and the couplings
of the theory such that $C$ becomes effectively a
decreasing function of $\tau$. In the case such a scheme
exists, it will certainly provide a strong constraint on the
structure of the underlying theory.
\medskip

Let us conclude this Section by noting that the signature
change model of compactification might possibly provide a
new mechanism for creating density fluctuations. In the
FRW ansatz (\ref{2.1}) we have neglected small
inhomogeneities. However, if ${\bf S}^3$ is slightly
distorted, one would expect local versions of our
dynamical variables to oscillate in different points with
slightly different amplitudes and phases. This should in turn
have some imprint on the resulting effective metric
(and the matter density in the case additional fields
are present). To which
extent such perturbations are wiped out by the expansion
is yet another interesting question to be pursued.

\section{Concluding remarks}
\setcounter{equation}{0}

In Sect.4 we have worked out the effective metric for the case
that the mixed trajectory oscillates along the lower branch of
the $W=0$ curve. This led to the compactification of ${\bf S}^6$.
One may of course reverse the situation and consider a mixed
trajectory moving near the upper branch, thus describing
compactification of ${\bf S}^3$ (i.e.
$a_1 \rightarrow \ell_\Lambda \sqrt{3}$, while
$a_2 \rightarrow \infty$). Neglecting -- in analogy with the
previous case -- the curvature contributions of ${\bf S}^6$,
amounts to omit the $u^{3/2}$-term in the potential (\ref{2.10}).
After a change of variables of the type (\ref{2.12}), namely
\begin{equation}
u=X^{1/3}, \qquad v=Y^{5/6}
\label{6.1}
\end{equation}
(cf. equation (\ref{4.3})), we arrive at a potential
\begin{equation}
\overline{W} =
   \frac{5}{18} \Big( -\xi^{-2/3} + 2\, \xi^{-1/4} \Big)
\label{6.2}
\end{equation}
with $\xi \equiv X/Y$ (cf. equation (\ref{4.4})). This is
very similar in structure to (\ref{4.5}), and without
having done the computation in detail, we expect that this
case yields a result analogous to (\ref{4.34}). Thus, in total,
letting the initial (Euclidean) trajectory starting from the
origin, there are three generic possibilities for the
long-time evolution: compactification of either
scale factor and escape into the Lorentzian region. In this
last case, both scale factors expand exponentially. In addition,
there are two isolated solutions  (namely if the initial
trajectory coincides with one of the instantons). Assuming
a reasonable measure on the set of initial conditions,
we expect the three generic cases to occur with comparable
finite probabilities. Computing such probabilities in detail
(possibly on the basis of a path integral formulation
suggesting a suitable measure)
would reveal whether a four-dimensional effective space-time is
favoured over a seven-dimensional one.
\medskip

Let us finally comment on the question of the dimensions
that may be put into a signature change model of
compactification from the outset. A natural generalization
of the version we were dealing with in this paper is to consider
a FRW model based on ${\bf S}^m \times {\bf S}^n$
as spatial sections. The most important thing to notice
in this context is that a ''branch'' of the
zero potential curve is lost if the curvature of some
factor space is zero. In a
${\bf S} \times {\bf S}^n$ (or, more general, a
${\bf T}^m \times {\bf S}^n$ or ${\bf R}^m \times {\bf S}^n$)
model with $n>1$, only ${\bf S}^n$
can compactify by signature change. In this sense, curvature
acts as an ''attractive force'' upon the corresponding
scale factors. One might try to formulate a statement like:
Signature change compactifies and stabilizes only internal spaces
with non-vanishing Ricci-curvature.
\medskip

Another observation in this context is that -- when trying to
express the action in terms of ''lightlike'' variables
along the lines of (\ref{2.9}) -- some ugly numerics
appears. In the general case, the potential is of the structure
\begin{equation}
W = \sum_{i=1}^3 c_i\, u^{\alpha_i} v^{\beta_i},
\label{6.3}
\end{equation}
up to transformations of the type (\ref{2.12}).
Among all combinations $m \leq n<10$ we find only for
$(m,n) = (2,8),(3,6),(5,5),(6,10)$ and $(7,8)$ that
$\alpha_i$ and $\beta_i$ are rational numbers. As an example,
for $(m,n)=(2,3)$, a simplification of the potential similar
to (\ref{2.9}) produces the exponents
\begin{eqnarray}
\alpha_1 = 0,\qquad
\alpha_2 = 1 + 2 \sqrt{\frac{2}{3}},\qquad
\alpha_3 = 1 + \sqrt{\frac{3}{2}},
\nonumber\\
\beta_1 = \frac{1}{2} + \sqrt{\frac{3}{2}},\qquad
\beta_2 = 0,\qquad
\beta_3 = \frac{1}{2}\,\alpha_3.\label{6.4}\\
\nonumber
\end{eqnarray}
In general, one encounters the square roots of
$m(m+n-1)/n$ and $n(m+n-1)/m$.
Nevertheless, the overall structure of the potential is
comparable to (\ref{2.10}), and we expect analogous effects to
arise here as well. Whether the ''beautiful'' cases
mentioned above are distinguished from the others, and
what we can learn from structures like (\ref{6.4}) about
the possibility and physics of signature change induced
compactification in various dimensions are again problems
that deserve further study.
\\
\\
\\

\bigskip
\bigskip
\bigskip

\noindent
{\Large {\bf Figure captions:}}
\\
\\
\\
{\large {\bf Fig.1}}\\
A typical mixed trajectory $(u(t),v(t))$ in minisuperspace
is shown. The dashed curve is $W=0$ (with its two
''branches''). The initial condition
of the trajectory near the origin is
$u^{-1/2} v \rightarrow 0.5$. The gauge condition defining $t$
is $N=1$, and the evolution has been performed numerically.
The first intersection of the trajectory with the
zero potental curve has actually a horizontal tangent
($\dot{v}=0$) -- this fact is suppressed by the low
resolution of the graphics. The long-time behavior,
which is essential for our purposes, is illustrated: the
trajectory remains near the lower branch of the dashed
curve, which implies compactification of ${\bf S}^6$.
The reverse situation, i.e. a trajectory oscillating
around the upper branch in much the same way,
is possible as well.
\\
\\
{\large {\bf Fig.2}}\\
This shows the graphs of the functions $a_1(t)$ and $a_2(t)$
corresponding to the trajectory displayed in Fig.1. The
unit on the vertical axis is $\ell_\Lambda$. The plot
demonstrates that $a_1$ behaves effectively linear
in $t$ when the universe is already large, and that
$a_2$ converges to its limiting value
$\ell_{\Lambda} \sqrt{15}$.
\\
\\
{\large {\bf Fig.3}}\\
The graph of $a_2(t)$ is displayed using a better resolution
than in Fig.2. It shows that $a_2$ performs
damped oscillations around its limit.
\\
\\
{\large {\bf Fig.4}}\\
This plot shows a numerical solution of the dynamical
system (\ref{4.10}-\ref{4.12}). The dashed line
is $\sigma(t)$, the solid line represents $5\,(z(t)-1/2)$
(the magnification factor of 5 has been introduced
for convenience). The initial
conditions have been chosen as
$z(0)=0.6$ and $\sigma(0)=0.31$, the gauge condition
defining $t$ is ${\cal N} = (2/9) (3/\Lambda)^{1/2}$.
The zigzag limit of $\sigma$ as well as the damped
nature of the oscillations are well illustrated. The
Lorentzian periods are those for which the solid
graph has positive values (or, equivalently,
during which the dashed graph increases).

\end{document}